\documentclass[letterpaper,11pt]{article}
\usepackage[T1]{fontenc}

\usepackage{amsthm}  
\usepackage{amsmath, amssymb}
\usepackage{color}
\usepackage{fullpage}
\usepackage{hyperref}
\usepackage[noend]{algorithmic}
\usepackage{thmtools, thm-restate}
\usepackage{enumerate}
\usepackage{thm-restate}
\usepackage{graphicx}
\usepackage{amsopn}
\usepackage{caption}
\usepackage{hyperref}
\usepackage{subcaption}
\usepackage{csquotes}
\usepackage{enumitem,linegoal}
\usepackage{comment}
\usepackage{float}
\usepackage{complexity}

\usepackage{ctable}					

\usepackage{mathtools}
\usepackage[flushleft]{threeparttable}
\usepackage[normalem]{ulem}

\usepackage{footnote}
\makesavenoteenv{tabular}
\makesavenoteenv{table}

\usepackage[margin=1in]{geometry}

 \newtheorem{theorem}{Theorem}

\theoremstyle{definition}

\makeatletter
\newif\ifqed
\def\GrabProofArgument[#1]{ #1: \egroup\ignorespaces}
\def\proof{\noindent\textbf\bgroup Proof%
	\@ifnextchar[{\GrabProofArgument}{. \egroup\ignorespaces}\global\qedtrue}

\def\qedhere{\ifmmode\tag*{\qedsign}\else\hspace*{\fill}\qedsign\medskip\fi\global\qedfalse}
\def\qedsign{$\Box$}
\makeatother

\newcounter{proccnt}

\newcommand{\konote}[1]{}

\title{A Simpler Approach to Linear Programming}
\author{Jean-Louis Lassez\\
IBM Research Center (Retired)\\
jlassez@gmail.com}
\date{}

\usepackage{fancyhdr}

\makeatletter
\renewcommand{\section}{\@startsection
{section}
{1}
{0mm}
{-1.2\baselineskip}
{0.01\baselineskip}
{\Large\bfseries\color{black}}} 

\begin{document}
\renewcommand{\theenumi}{(\roman{enumi})}
\renewcommand{\labelenumi}{\theenumi.}
\sloppy

\maketitle

\setlength{\parskip}{0pt}
\begin{abstract}\noindent
Dantzig and Eaves claimed that fundamental duality theorems of linear programming were a trivial consequence of Fourier elimination. Another property of Fourier elimination is considered here, regarding the existence of implicit equalities rather than solvability. This leads to a different interpretation of duality theory which allows us to use Gaussian elimination to decide solvability of systems of linear inequalities, for bounded systems.

\end{abstract}

\setlength{\parskip}{9pt}
\setlength{\parindent}{0em}

\section{Introduction: Another Facet of Fourier Elimination}
Dantzig and Eaves~\cite{3} make an important statement regarding Fourier elimination:

``From it one can derive easily, by trivial algebraic manipulations, the fundamental theorem of linear programming, Farkas lemma, the various theorems of the alternatives and the well known Motzkin Transportation theorem.''

Because Fourier algorithm is about elimination and Tarski showed that any theorem in elementary algebra and geometry can be obtained via elimination, this remark about the capacity of Fourier elimination to provide proofs of major theorems is not accidental. 
What is more striking is the use of the word ``trivial''.
The essence of Dantzig's and Eaves remark is that theories of duality and associated theorems of the alternative are based on the fact that Fourier's algorithm generates a contradiction $0 \leq -1$ when the primal is not solvable. In~\cite{10} it was shown that when Fourier algorithm generates a tautology $0 \leq 0$ it tells us that the primal has implicit equalities. This gives us a parallel theory of duality with associated theorems of the alternative, also obtained by trivial operations. In particular the primal is solvable or solvable at infinity if and only if  the elementary dual  has implicit equalities.  And the strong duality theorem has a new interpretation which leads to an algorithm to decide if a bounded set is solvable. The algorithm uses Gaussian elimination until the last step . Then it uses a trivial case of Fourier elimination. It has three interesting aspects, one from a theorem proving point of view, the algorithm comes very naturally and its correctness is straightforward, in the same line as Dantzig and Eave's statement. The second point is that it solves an important problem, known as Smale's problem number nine about finding a strongly polynomial algorithm. The third point is that one could have found a solution for Smale's problem number nine with a polynomial of arbitrary high degree, making it unsuitable for practical applications. Here the complexity is based on Gaussian elimination, for which there are many efficient implementations. Consequently, this new algorithm may have significant practical use. A topic we are exploring.
Now simplicity is in the eye of the beholder. Here we claim simplicity as all the operations are, as noted by Dantzig and Eaves ``trivial'' manipulations whose correctness is an immediate consequence of Fourier algorithm. The exception is the use of Gaussian elimination, which is not a trivial operation, but can be considered a simple one.

\section{Elementary Duality}
A trivial remark gives us a different view of duality. We need a definition first:
A constraint $L \leq r$  is an implicit equality if and only if the constraint $-L \leq -r$  holds, that is if and only if adding the two constraints gives us the constraint  $[0] \leq 0$,
where $[0]$ denotes the constraint with all coefficients set to $0$.  
This obvious remark can be generalized easily, as was shown in~\cite{10}.

The constraints $\{ L_i   \leq  r_i  \}$ are  implicit equalities if and only if there exists a set of positive coefficients $\{\lambda_i\}$, called multipliers   giving rise to  linear combinations  
   $\sum  \lambda_i L_i =[0]$   and $\sum  \lambda_i r_i =0$. 

It might not immediately appear so but, as we will see, this is in fact an obvious statement of a theorem of the alternative.   

The elementary dual of a set $S$ of constraints
$AX \leq b  \cup \{x_j  \geq 0 \}$ is $A^T \Lambda \geq 0 \cup \{-\sum  \lambda_i r_i  \geq 0\} \cup \{\lambda_i \geq 0\}$ where $A^T$ represents the transpose of A. 
And the $r_i$ are the right hand side constants that form the vector $b$.
The constraint $-\sum  \lambda_i r_i  \geq 0$ is called the extension.

Once we remark that   a constraint  $L_i  \leq  r_i$  can be written equivalently as   $L_i+s_i - r_i = 0$
where  $s_i$ is a non negative variable called a slack variable, the following becomes obvious:

The coefficients $x_j$ and $s_i$  of a solution of $S$ together with the coefficient 1 for the extension form a set of multipliers for a set of implicit equalities in the elementary dual.

In other words, $S$ has a solution if and only if the extension in the elementary dual is an implicit equality. 

Equivalently S has no solution if and only if the extension is not an implicit equality, that is the elementary dual has a solution such that the extension is $>0$, this is Farkas lemma. 

This is because the solution in the elementary dual corresponds to multipliers in the primal giving us $0 \leq -1$

In fact we have a more complete version of the lemma, which tells us that if the elementary dual has a solution, the primal is unsolvable or has an implicit equality.

This implies also that if Farkas lemma is a trivial consequence of Fourier algorithm, the previous remark tells us that Farkas lemma implies the correctness of Fourier algorithm. That is because an unsolvable set remains unsolvable when Fourier elimination is applied. 

A non solvable set may have implicit equalities, with an appropriate set of multipliers. For instance it may have a solvable subset.  But also as it is not solvable, Fourier elimination will produce a constraint $[0] \leq  -1$. If we add the tautology $0 \leq 1$, we have multipliers that produce $[0] \leq 0$.

A set $AX \leq b$  is said to have a \emph{solution at infinity} if and only if the set $AX \leq 0$ has a solution different from the origin.
A set without solutions at infinity is called \emph{bounded}.
Solutions at infinity correspond to implicit equalities in the elementary dual independent of the extension.

\medskip\noindent\textbf{Example:}

\noindent 
The set

\[\begin{array}{rrrr}
-x&+y &\leq& 2 \\
 x&-y  &\leq& -1 \\
-x&     &\leq& 0 \\
  & -y  &\leq& 0 
\end{array}\]

\noindent 
Can be rewritten as:
\[\begin{array}{rrrrrr}
-x&+y& + s_1&     - 2 &=&0 \\
 x&- y & + s_2 & + 1 &=&0 \\
-x&  & &             &\leq& 0 \\
  &  -y & &             &\leq& 0 
\end{array}\]

\[\begin{array}{rrrrrl}
0 &  1  &  1 &0 & 1 &                \text{solution} \\
1  & 1 &   0 &0&  0 &               \text{solution at infinity}
\end{array}\]

\noindent
It has the elementary dual

\[\begin{array}{rrrr cc}
                      &&&&                     \text{solution}         &       \text{solution at infinity} \\
                      &&&&                     \text{multipliers}          &      \text{multipliers} \\
  -\lambda_1 & +\lambda_2 &\geq& 0    &            0 \; x                   &            1 \; x \\
   \lambda_1 & -\lambda_2  &\geq& 0   &              1 \; y                 &              1 \; y \\
-2\lambda_1 & +\lambda_2 &\geq& 0     &            1 \; 1                   &            0 \; 0 \\
   \lambda_1 &         &\geq& 0       &         \; 1 \; s_1                   &          \; 0 \; s_1 \\
         & \lambda_2 &\geq& 0         &        \; 0 \; s_2                     &         \; 0 \; s_2 
\end{array}\]

If the right hand side coefficients 2 and -1 are respectively replaced by -2 and 1 the extension is not an implicit equality as $\lambda_1 = \lambda_2 =1$ make the extension $>0$

We can reword the theorem of the alternative in the following way:

\begin{theorem}
\textbf{Solutions and solutions at infinity in the primal are multipliers in the elementary dual.}
\end{theorem}

The fundamental theorem on strong duality states that  

For a set of constraints
$AX \leq b  \cup \{x_j  \geq 0 \}$ the maximum of $\sum  c_i x_i$ with $c$ being the vector of the $c_i$, and $b$ a vector of coefficients denoted $r_j$ is obtained when the following system is solvable:

\begin{align*}
& AX \leq b  \cup \{x_j  \geq 0\} \\
& A^T \Lambda \geq c \\
& \sum  \lambda_i r_i =\sum c_i x_i  \cup \{\lambda_i \geq 0\}. \\
\end{align*}

This system can be written equivalently where the left hand side of the constraints are those of the elementary dual.
We have added to the primal a new set of constraints called the strong elementary dual .

\begin{align*}
& AX \leq b  \cup \{x_j  \geq 0\}  \\
& A^T \Lambda \geq c \\
& -\sum  \lambda_i r_i  \geq  -\sum c_j x_j  \cup \{\lambda_i \geq 0\}.
\end{align*}

Let us also call this last constraint the extension

Because a solution to the system will give multipliers such that  $\sum c_j x_j - \sum c_j x_j = 0$, we have that the maximum is reached for the extension to be an implicit equality with multiplier equal to 1.

\medskip\noindent\textbf{Example:}

\[\begin{array}{rrrr}
 Max(x+y)&\\
x&+y &\leq& 4 \\
x&-y &\leq& 0 \\
-x&+y &\leq& 0 \\
-x& &\leq& -1 \\
  &   -y &\leq& -1 \\
-x& &\leq& 0 \\
   &  -y &\leq& 0 \\
\end{array}\]     

In the standard duality we add the constraints:

\[\begin{array}{rrrrrrl}
\lambda_1 &+  \lambda_2 &-  \lambda_3 &-  \lambda_4   &   &\geq& 1 \\
\lambda_1 &- \lambda_2 &+  \lambda_3  & &   - \lambda_5   &\geq& 1 \\
4\lambda_1 &- &&\lambda_4 &- \lambda_5&=&x+y
\end{array}\]   


Together with sign constraints

Let us write it this way:
\[\begin{array}{rrrrrr lr ll}
                  &&&&&&                       && \text{Multipliers}  \\
\lambda_1 &+ \lambda_2   &-\lambda_3  &-\lambda_4  &      &\geq &1        &&   x \\
\lambda_1 &- \lambda_2  &+  \lambda_3 &      &- \lambda_5   &\geq &1        &&   y \\
\\
 4\lambda_1 & &&                – \lambda_4 &– \lambda_5  &\geq & x+y    &&  0 \\
-4\lambda_1 & &&         + \lambda_4 &+ \lambda_5  &\geq  &-x-y           &&   1 \\ 
\end{array}\] 

Together with sign constraints.

If this last constraint is an implicit equality, this formulation is equivalent to the standard one when there is a solution. And when there is a solution it satisfies the last constraint as an equality as the multipliers give us $[0]\geq 0$

From \cite{10} we know it means the last constraint is an implicit equality.

An advantage of this formulation is that it is more compact than the standard one. Because the information for the solutions in the primal is contained as multipliers in the strong elementary dual.

\section{Multipliers and Primal cone}
The following properties of multipliers are as immediate as they are fundamental.
It is however  a delicate situation because we have to be very careful when we use the words solution or solution at infinity, multipliers or implicit equalities. 
The following statements are straightforward when we use them in the right context, and we realize that they follow directly from the two properties of Fourier algorithm regarding solvability and implicit equalities. 

If the primal is solvable it has a set of multipliers with the multiplier for the extension equal to 1. However multipliers can be multiplied by an arbitrary positive scalar. So the requirement that the extension' s multiplier be equal to 1 can be replaced by the requirement that the multiplier be positive. 

Let $AX \leq b$ together with non negativity constraints for the variables  be the primal. Let z be the name of a new variable. The primal cone is defined by adding the column
 $-bz$ to $AX$, setting the right hand side constants to 0,  with the same non negativity constraints for the variables  as well as for the variable z.  

If the primal is bounded, then the primal cone is reduced to the origin if and only if the primal is unsolvable. 

Indeed if the primal cone has a solution X different from the origin, z is $>0$ as the primal is bounded. 
And the primal is solvable by dividing the X solution with z. 
And if the primal has a solution, obviously the primal cone has one different from the origin.
This is important as we know that to determine solvability we can restrict ourselves to the bounded case~\cite{1}.

And also the immediate:

If the primal cone has multipliers the primal is either  non solvable or non full dimensional.

This depends on whether the multiplier for $-z \leq 0$  is zero or not. If it is positive, it tells us that when all variables are eliminated we are left with a contradiction in the primal. 

If the primal is bounded and unsolvable, its elementary dual has no implicit equalities. Equivalently all multipliers in the elementary dual are equal to zero. If the primal is  bounded and solvable, all sets of multipliers have a positive multiplier for the extension. If the primal is solvable and full dimensional, all the constraints in the elementary dual are implicit equalities and the elementary dual is reduced to the origin. Conversely if the elementary dual is reduced to the origin, the primal is solvable and full dimensional. So an algorithm to decide if a cone is reduced to the origin can be used to decide if a set, bounded or not, is full dimensional solvable. Or more directly  a set is solvable  full dimensional if and only if its primal cone is full dimensional.

Assuming the input to be bounded, it may be solvable full dimensional, solvable with implicit equalities or not solvable.
In the first case its primal cone is full dimensional and its dual cone is reduced to the origin. In the second case its primal cone has implicit equalities but is not reduced to the origin. In the third case, its primal cone is reduced to the origin and its dual is a full dimensional cone.

\section{Parasite Multipliers}
Consider the set:

\[
  x+y \leq 0, 2x-y \leq 0, -x+2y \leq 0
\]

If we eliminate x using Fourier elimination, we obtain:

$y \leq 0$ and $3y \leq 0$, the multipliers for this set are 0 and 0 as it has no implicit equality.

If we eliminate $x$ by Gaussian elimination using the first constraint, setting $x=-y$ we obtain:

$-3y \leq 0$ and $y \leq 0$.  This set admits the multipliers 1 and 3. 
We call these parasite multipliers.

However, as we see now, if the initial set has multipliers, eliminating a variable by Gaussian elimination  setting a symbol $=$ instead of  $\geq$  or  $\leq$    in a main constraint with a positive multiplier gives a new set which also has multipliers that are derived from those of the initial set. We call these legitimate multipliers. It may also generate parasite multipliers. When the new set has multipliers we can decide if they are legitimate or parasite.
From the legitimate ones we  can retrieve  multipliers for the initial set. The parasite multipliers tell us that the constraint used to eliminate a variable in the initial set is redundant, as a positive linear combination of some other constraints in the set. It is straightforward, but there are a lot of details we have to consider.  And of course if we eliminate a variable by such Gaussian elimination using a constraint that is not an implicit equality, we may generate an unsolvable set.

Formally:
We consider a cone with  $\leq$  symbols. This cone can be a primal cone. 
We choose a variable $x_0$, the other variables are labelled $x_v$  the index v positive and a main constraint $a_0 x_0 + L_0  \leq 0$.  The requirements are that  $a_0$  be different from 0  and that $L_0$ is not reduced to $[0]$.  Otherwise we are in a trivial case where all main constraints are $[0] \leq 0$ and if  a main constraint is such that $L_0$ is reduced to $[0]$, then it can be eliminated as redundant or the variable set to 0. In such cases multipliers will be adapted to the situation.

The notation is simplified if we first  scale the coefficients of the variable $x_0$ in such a way that they become equal to 1, -1, or remain 0.
If we have a constraint $a x_0 + L \leq 0$  with a multiplier $\mu$, and $a$ is positive, it becomes
$x+1/(a)  L \leq 0$  or $-x+1/(-a) L \leq 0$ if $a$ is negative.  
The multiplier $\mu$ becomes $a\mu$  or $-a\mu$. 
The multiplier for the $-x_0  \leq 0$ will not be modified as we see below.

The constraints are classified in the following way

\[\begin{array}{rrrr c}
                          &&&&                  \text{Multipliers} \\
  a_0 x_0 &+ L_0 &\leq 0                   &&            \mu_0      \\    
  a_i x_0 &+ L_i  & \leq 0                  &&             \mu_i  \\
 -a_j x_0 &+ L_j  & \leq 0                 &&              \mu_j \\
          &  L_k  & \leq 0                  &&             \mu_k  \\
-x_0 &  & \leq 0                           &&           s_0  \\
    &   -x_p   &\leq 0                    &&              s_p      \\       
\end{array}\] 

The sets of indices $i$, $j$, $k$, are distinct sets and do not contain the index 0.  
And all indices p are different from 0.
All the $a_0, a_i, a_j$ are positive.

\[\begin{array}{rrrr l}
 x_0 &+1/a_0 L_0  & \leq 0               &&             a_0\mu_0 \\
 x_0 &+ 1/a_i L_i  & \leq 0              &&             a_i\mu_i \\
 -x_0 &+ 1/a_j L_j & \leq 0              &&             a_j\mu_j \\
     &     L_k   & \leq 0              &&                     \mu_k \\
-x_0 &   & \leq 0                        &&                    s_0 \\
    &   -x_p  & \leq 0                   &&                  s_p \\             
\end{array}\]

We eliminate $x_0 = -1/a_0 L_0$

And obtain a new set of constraints with multipliers

\[\begin{array}{rrrr c}
                   &&&&                              \text{Multipliers} \\
  -1/a_0 L_0&+1/a_i L_i   &\leq 0                      &&     a_i \mu_i  \\
   1/a_0 L_0&+ 1/a_j L_j  &\leq 0                      &&     a_j \mu_j  \\
          L_k&             &  \leq 0                   &&            \mu_k  \\
   1/a_0 L_0  &             &\leq 0                    &&           s_0  \\
       & -x_p   &\leq 0                                 &&            s_p   \\  
\end{array}\] 

It is straightforward to show that the multipliers of the initial set are transferred in the way described. The important point is that $s_0$ the slack multiplier of a sign constraint becomes the multiplier of a main constraint.
If this multiplier is positive, the main constraint is an implicit equality.
 
The process is reversible as $a_0 \mu_0 = -\sum a_i\mu_i + \sum a_j \mu_j + s_0$
Parasite multipliers will fail this reversal.

\section{Algorithm to test if a bounded system is solvable}
We know that a bounded system is solvable if and only if its primal cone is not reduced to the origin. 
So in a first step we have as input a bounded system and we go to its primal cone.

A constraint $\sum x_i  \leq 2$  is added to the primal cone. 
Call this system the primal.
We then use the new version of the strong duality theorem of linear programming to compute $\max(\sum  x_i)$. 
We illustrate the algorithm with an example.
The new system consists of two parts, the primal and its associated strong elementary dual.

\[\begin{array}{rrrr}
Max (x+y)  \\
x+y &\leq 2  \\
x-2y &\leq 0  \\
x-y &\leq 0  \\
x-3y &\leq 0  \\
-x &\leq 0  \\
-y &\leq 0  \\
\end{array}\]

\[\begin{array}{rrrrl r c}
                       &&&&&&              \text{Multipliers} \\
\lambda_1 &+ \lambda_2 &+ \lambda_3 &+ \lambda_4  &\geq 1         &&        x  \\
\lambda_1 &-2\lambda_2 &-  \lambda_3 &-3\lambda_4  &\geq 1      &&        y     \\    
-2\lambda_1 &&&  &\geq -x-y=-\sigma= 0 \text{ or} -2                    &&    1   \\
\lambda_1 &&& &\geq 0                                            &&    s_1   \\
     & \lambda_2 && &\geq 0                                      &&    s_2   \\
    &&       \lambda_3 & &\geq 0                                 &&    s_3   \\
     &&&           \lambda_4 &\geq 0                            &&    s_4   \\
\end{array}\]

There is always a solution either the origin  with $\sum x_i=0$   or a solution such that the maximum is reached for $\sum x_i=2$ 
Assume the maximum is 2, set $\sigma=2$.  The new system is such that  $\sum x_i  \leq 2$  is an implicit equality, it is solvable. The multiplier for the extension is equal to 1, and the strong elementary dual is solvable.  
The new system has a remarkable property. 
As $\lambda_1$ is equal to 1, it is solvable for all coefficients of the other $\lambda$'s, including the situation where they are all 0. And the strong elementary dual is equal to  the elementary dual of the original cone.  

Now we eliminate a $\lambda_i$   different from  $\lambda_1$  using a main  constraint in the strong elementary dual, different from the extension, where the inequality is replaced by an equality.

\[
1-\lambda_1 - \lambda_2  - \lambda_4 =\lambda_3
\]

\[\begin{array}{rrrl rl }
        &&&&&                          \text{Multipliers} \\
 2\lambda_1 &– \lambda_2 &- 2\lambda_4   & \geq 2       &&           u>0   \\        
-2\lambda_1 &            &     & \geq -x-y= -2          &&           1    \\
-\lambda_1 &-   \lambda_2  &-  \lambda_4  &\geq -1      &&           s_3>0   \\
 \lambda_1 & &  &\geq 0                                                  &&                         s_1  = 0   \\
     & \lambda_2 &   &\geq 0                                            &&                         s_2   \\
     &  &         \lambda_4   &\geq 0                                 &&                          s_4   \\
\end{array}\] 

We are in fact working on the elementary dual of the original cone, and this operation can set to 0 several if not all $\lambda$'s, as parasite multipliers can be created. But that does not affect the solvability of the system as $\lambda_1$ remains equal to 1. And the extension remains an implicit equality. What happens is that a solvable system becomes even more solvable. 
If the original cone is reduced to the origin, we have $\lambda_1=0$, having set  $\sigma=2$ renders the new system unsolvable. Either 
the strong elementary dual is not solvable, or the extension is not an implicit equality. Eliminating a variable as was done preserves unsolvability. If the strong elementary dual is solvable and eliminating a variable still gives a solvable set, all solutions are such that $\lambda_1=0$. We cannot have the extension be an implicit equality, which would imply $\lambda_1=1$.

We eliminate in the same way all the other variables but the variables $\lambda_1$.

\[\begin{array}{rll l}
             &        &&                       \text{Multipliers} \\
-2\lambda_1 &\geq -2                      &&                             1>0   \\
-2\lambda_1 &\geq -x-y=-2                  &&                       1/2>0   \\
  3\lambda_1 &\geq 3                    &&                               1/3>0   \\                       
   \lambda_1 &\geq 0                     &&                               s_1=0   \\
\end{array}\] 

This last system is solvable for $x+y=2$,  and the extension is still an implicit equality. 
And we find that $\lambda_1$ is equal to 1.

\medskip\noindent\textbf{Example:}
Input cone is $x+y \leq 0$, $-x-y \leq 0$, $x$, $y$, non negative. 
It is reduced to the origin.
The process gives us the two constraints $\lambda_1 \leq 1$ and  $\lambda_1 \geq 0$. While the only two possible solutions are $\lambda_1=1$ or 
$\lambda_1=0$, and they are mutually exclusive.
So we find $\lambda_1=1$ if and only if the original bounded system is not solvable.

The references \cite{1,2,3} and \cite{10} are used to establish the results in this paper. 
The other references~\cite{4,5,6,7,8,9,11,12} are to provide background information on various aspects of Fourier elimination.

\bibliographystyle{alpha}
\bibliography{paper}

\begin{thebibliography}{BLLM99}

\bibitem[Ach84]{1}
S~Achmanov.
\newblock Programmation lin{\'e}aire traduction fran{\c{c}}aise.
\newblock {\em MIR, Moscou}, 1984.

\bibitem[BLLM99]{5}
Alexander Brodsky, Catherine Lassez, Jean-Louis Lassez, and Michael~J Maher.
\newblock Separability of polyhedra for optimal filtering of spatial and
  constraint data.
\newblock {\em Journal of Automated Reasoning}, 23(1):83--104, 1999.

\bibitem[CL03]{4}
Vijay Chandru and Jean-Louis Lassez.
\newblock Qualitative theorem proving in linear constraints.
\newblock In {\em Verification: Theory and Practice}, pages 395--406. Springer,
  2003.

\bibitem[DE73]{3}
George~B. Dantzig and B.~C. Eaves.
\newblock Fourier-motzkin elimination and its dual.
\newblock {\em Journal of Combinatorial Theory Ser. A}, page 255, 1973.

\bibitem[Fou27]{2}
JBJ Fourier.
\newblock Analysis of the work of the royal academy of sciences during the year
  1824.
\newblock {\em Mathematic Part}, 1827.
\newblock Partial English translation by D.A. Kohler, ``Translation of a Report
  by Fourier on his work on Linear Inequalities,'' Opsearch 10.

\bibitem[HJCL91]{6}
T.~Huynh, L.~Joskowicz, and Jean-Louis~Lassez Catherine~Lassez.
\newblock Practical tools to reason about linear constraints.
\newblock {\em Fundamenta Informaticae}, 15(3-4), 1991.

\bibitem[HN94]{8}
Dorit~S Hochbaum and Joseph Naor.
\newblock Simple and fast algorithms for linear and integer programs with two
  variables per inequality.
\newblock {\em SIAM Journal on Computing}, 23(6):1179--1192, 1994.

\bibitem[LHM92]{12}
Jean-Louis Lassez, T.~Huynh, and K.~McAloon.
\newblock Simplification and elimination of redundant linear arithmetic
  constraints.
\newblock {\em Constraint Logic Programming, Selected Research}, 1992.

\bibitem[LM92a]{10}
Jean-Louis Lassez and Michael~J Maher.
\newblock On fourier's algorithm for linear constraints.
\newblock {\em Journal of Automated Reasoning}, 9(1), 1992.

\bibitem[LM92b]{11}
Jean-Louis Lassez and K.~McAloon.
\newblock A canonical form for generalized linear constraints.
\newblock {\em Journal of Symbolic Computation}, 1992.

\bibitem[Sma98]{7}
Steve Smale.
\newblock Mathematical problems for the next century.
\newblock {\em The mathematical intelligencer}, 20(2):7--15, 1998.

\bibitem[YL82]{9}
Boris Yamnitsky and Leonid~A Levin.
\newblock An old linear programming algorithm runs in polynomial time.
\newblock In {\em 23rd Annual Symposium on Foundations of Computer Science
  (sfcs 1982)}, pages 327--328. IEEE, 1982.

\end{thebibliography}

\end{document}